\documentclass[compsoc, conference, a4paper, 10pt, times]{IEEEtran}

\usepackage{cite}
\usepackage{amsmath,amssymb,amsfonts}
\usepackage{algorithmic}
\usepackage{graphicx}
\usepackage{textcomp}
\usepackage{xcolor}
\usepackage{booktabs}
\usepackage[hidelinks]{hyperref}
\usepackage{tabularx,booktabs}

\begin{document}

\title{Architecture of Data Anomaly Detection-Enhanced Decentralized Expert System for Early-Stage Alzheimer’s Disease Prediction}


\author{
  \IEEEauthorblockN{1\textsuperscript{st} 
    Stefan Kambiz Behfar
  }
  \IEEEauthorblockA{
    \textit{Department of Information Systems, Geneva School of Business Administration, Switzerland} \\
    \textit{Department of Computer Science and Technology, University of Cambridge, United Kingdom} \\
    stefan-kambiz.behfar@hesge.ch
  }
  \and
  \IEEEauthorblockN{2\textsuperscript{nd} 
    Qumars Behfar
  }
  \IEEEauthorblockA{
    \textit{Department of Neurology, Faculty of Medicine and University Hospital Cologne, University of Cologne, Germany} \\
    qumars.behfar@uk-koeln.de
  }
  \and
  \IEEEauthorblockN{3\textsuperscript{rd} 
    Marzie Hosseinpour
  }
  \IEEEauthorblockA{
    \textit{Institute of Business Informatics, Faculty of Business Economics, Hasselt University, Belgium} \\
    marzie.hosseinpour@uhasselt.be
  }
}

\maketitle

\begin{abstract}
    Alzheimer’s Disease, a global health challenge, necessitates early and precise detection to enhance patient outcomes. Magnetic Resonance Imaging (MRI) offers significant diagnostic potential, but its effective analysis remains a formidable task. This study presents a groundbreaking decentralized expert system, which ingeniously merges blockchain technology with Artificial Intelligence (AI), integrating robust anomaly detection for patient-submitted data.
    Traditional diagnostic methodologies often result in delayed and imprecise predictions, particularly in the disease's early stages. Centralized data repositories struggle to manage the immense volumes of MRI data, alongside persistent privacy concerns that impede collaborative efforts.
    Our innovative solution harnesses the power of decentralization to safeguard both data integrity and patient privacy, facilitated by blockchain technology. It not only emphasizes AI-driven MRI analysis but goes a step further by incorporating an architecure of a sophisticated data anomaly detection mechanisms. These mechanisms are designed to scrutinize patient-contributed data for various issues, including data quality problems and atypical findings within MRI images. 
    Performing an exhaustive check of the correctness and quality of MRI images and biological information directly on-chain is not practical due to the computational complexity and cost constraints of blockchain platforms. Instead, such checks are typically performed off-chain, and the blockchain is used to record the results securely.
    This comprehensive approach empowers our decentralized app to provide more precise early-stage Alzheimer’s Disease prediction. Further, by merging the strengths of blockchain, AI, and anomaly detection, our aimed-system represents a pioneering step towards revolutionizing disease diagnostics.
\end{abstract}

\section{INTRODUCTION}
Magnetic resonance imaging (MRI) allows non-invasive examination of the brain. MRI quickly gained popularity in the medical field, eventually leading to a resourceful application, that today there are different types of MRI, including anatomical or structural (sMRI), diffusion MRI, Positron Emission Tomography MRI (PET-MRI), and functional MRI (fMRI), and the list is not exhaustive. The outcomes of MRI scans can be extremely versatile (see FSL website [1]). To achieve a final MRI image different sequences will be combined, where sequences are a type of image that reveals the specificity of the brain’s structure.

Artificial intelligence (AI) is the area of computer science focusing on creation of expert machines that engage on human-like intelligence (Russell and Norvig 2002 [2], Hope and Wild 1994 [3], Kasabov 1998 [4]).
The main source of an expert system is the obtained knowledge including a knowledge acquisition component that processes data and information and shapes them into rules. Expert systems have a large spectrum of application areas such as monitoring, prediction, classification, decision-making, planning etc. Importantly, medical diagnosis is one of the major applications of expert systems. Medical expert systems are to support the diagnostic process of physicians. This implies that a medical expert system employs knowledge about the diseases and compares with facts about the patients to suggest a diagnosis (Waterman 2009 [5]). Medical expert systems have been successfully implemented in diverse medical fields including neurology to improve the accuracy of diagnosis of neurological and neuropsychological disorders.

Alzheimer’s disease (AD) is one of the main neurodegenerative diseases and the leading cause of dementia. 
Research concerning AD evolves primarily around brain structural and functional analyses. For AD in particular, the functional analysis-derived network analysis is extremely helpful since it correlates different brain regions pointing to alternations of the neurological network and thus allowing quicker identification of the disease in its earlier stages. There are continuous demands to research in this domain. In fact, several studies have focused on the diagnosis of AD; Obi and Imainvan (2011) [6] developed a neuro-fuzzy model for the diagnosis of Alzheimer’s on the basis of neuropsychological tests including nine symptoms like memory loss, and difficulty in performing familiar tasks. Trambaiolli et al. (2011) [7] developed an AD diagnostic system based on a support vector machine which resulted in an accuracy of 79.9\% with 83.2\% sensitivity. 
Behfar et al. (2020) [8] used graph theory to reveal resting-state compensatory mechanisms in early-stages of AD. There are other and more recent studies that provide even better accuracies (Liu et al. 2023) [9]; however, all the studies suffer from a lack or shortage of longitudinal data on the patients, and to the best of our knowledge there has been very limited research that explores collection of such longitudinal data on AD patients via an application.

Our goal for this research is to design a decentralized expert system including a user-friendly web application to upload some biological information and MRI images
of the brain directly by the patients, keeping their data in a privacy-preserving manner, and propose an AI model to detect early-stage AD. This helps patients monitor their AD progression in time, also assists clinics who wish to use this software to monitor patients’ disease development.
In the first section, we discuss the research design and relevant questions, then provide our decentralized App solution in the next section, and provide the architecture, AI model, class diagram, and its implementation.

\section{RESEARCH DESIGN}
A three-dimensional image composed of various voxels can be either “white matter” which connects the neurons to each other and conducts impulses away from the soma, or the “grey matter” which is mostly made of neuron cell bodies, neuron somas which are the input unit of electrical signals sent within the central nervous system. Lastly, when examining an MRI image, there are hollow spaces, which are spaces filled with CSF and commonly referred to as “third tissue”. Brain parcellation or node definition is the name of the process that splits the brain into multiple different ROIs. \\
Prior to any analysis on MRI images, they are required to undergo a “cleaning process”, which is called prepro-cessing. Several factors can distort the outputs of an MRI scanning session and thus falsify the results. They are re-ferred to as noise and can have multiple sources, such as:
\begin{itemize}
    \item Minor movements of the person influencing the final MRI.
    \item Heartbeat, respiration, low-frequency oscillations.
    \item Complexity of the brain structure, tissues are not clearly distinguishable.
    \item Technical issues in the scanner, so-called scanner artifacts.
\end{itemize}
Once the preprocessing is performed via FSL Library (see FSL website [1]), the images can be analyzed depending on the type of MRI. We have created a web application (see https://github.com/stefankam/predprodalzheimer) which uses FSL and creates brain connectivity matrices using Octave (see GNU Octave [10]) with network modeling and pushes to the AI engine. 
This type of analysis is often performed on resting-state fMRI and describes brain functions by the connections and interactions between the highly interconnected brain regions (Sohn et al. 2017 [11]). By creating a brain connectivity matrix, these connections are contextualized and visualized for deeper understanding. 

\subsection{Related Work}
The development of decentralized expert systems for Alzheimer's Disease diagnosis and management is an emerging field, intersecting various domains such as AI, blockchain, and healthcare. This section explores existing work in these areas, emphasizing the uniqueness of our proposed system.

\subsubsection{AI in Alzheimer's Disease diagnosis}
\begin{itemize}
    \item \textbf{Description:} Research in this area focuses on using AI algorithms for early detection and prediction of Alzheimer's Disease, analyzing data from neuroimaging, genomics, and clinical assessments.
    \item \textbf{Limitations:} While AI models show promise, challenges remain in terms of data diversity, model interpretability, and generalization across different patient populations.
\end{itemize}

\subsubsection{Blockchain for secure medical data management}
\begin{itemize}
    \item \textbf{Description:} Blockchain technology is being explored for enhancing data security, patient privacy, and traceability in healthcare, with applications ranging from medical records management to drug traceability.
    \item \textbf{Limitations:} Blockchain solutions in healthcare often face scalability issues, and there's a need for balancing decentralization with regulatory compliance.
\end{itemize}

\subsubsection{Decentralized data sharing in healthcare}
\begin{itemize}
    \item \textbf{Description:} Initiatives like Health Level Seven International (HL7) and Fast Healthcare Interoperability Resources (FHIR) aim to standardize and facilitate secure data sharing in healthcare.
    \item \textbf{Limitations:} Despite these efforts, achieving interoperability and efficient data sharing while maintaining privacy and security remains a challenge.
\end{itemize}

\subsubsection{Ethical AI and data privacy}
\begin{itemize}
    \item \textbf{Description:} The ethical use of AI and the protection of patient data privacy are crucial considerations, especially in sensitive areas like Alzheimer's Disease.
    \item \textbf{Limitations:} Ensuring ethical AI practices and robust data privacy in decentralized systems is complex and requires continuous attention and adaptation to emerging technologies and regulations.
\end{itemize}

\textbf{Our Approach:}
Our proposed decentralized expert system for Alzheimer's Disease diagnosis and management integrates these diverse areas, offering a novel solution:
\begin{itemize}
    \item \textbf{Holistic AI-Driven Analysis:} Our system leverages advanced AI techniques for comprehensive analysis of medical data, ensuring accurate and early diagnosis of Alzheimer's Disease.
    \item \textbf{Blockchain-Enhanced Security and Privacy:} We utilize blockchain to secure patient data, ensuring integrity, traceability, and compliance with healthcare regulations.
    \item \textbf{Interoperable Data Sharing:} Our architecture supports standardized, secure, and efficient data sharing across different healthcare systems and stakeholders.
    \item \textbf{Ethical and Privacy-Centric Design:} We prioritize ethical AI practices and robust data privacy measures, aligning with the highest standards of patient care and data protection.
\end{itemize}

\subsection{Research Questions}
The final purpose of this implementation is to make longitudinal medical data linked to AD easily accessible to perform further disease prediction via a decentralized expert system. Using the application, fMRI data can be uploaded and processed. Thanks to the brain connectivity matrix as the principal output, the images can quickly be analyzed and interpreted. But,

\textbf{Research question 1:} What are the key factors influencing the accuracy and reliability of the decentralized expert system in diagnosing Alzheimer's Disease?

The main source of an expert system is the obtained knowledge including a knowledge acquisition component that processes data and information and shapes them into rules. Expert systems have a large spectrum of application areas such as monitoring, prediction, classification, decision-making, planning etc. Importantly, medical diagnosis is one of the major applications of expert systems. Medical expert systems are to support the diagnostic process of physicians. This implies that a medical expert system employs knowledge about the diseases and compares it with facts about the patients to suggest a diagnosis. In the case of AD,

\textbf{Research question 2:} How does the use of a decentralized expert system by healthcare professionals impact the diagnosis and treatment planning for Alzheimer's Disease?

 A decentralized expert system is a type that is built on a decentralized network of nodes, rather than being centrally controlled by a single entity. In this system, each node contains a subset of knowledge, and these nodes work together to make decisions. Decentralized expert systems have several advantages over traditional expert systems. They are more resilient and less vulnerable to a single point of failure, as there is no central point of control. Finally, they can be more transparent and secure, as each node can be verified and audited independently. But,

\textbf{Research question 3:} How does the performance of a decentralized expert system in diagnosing Alzheimer's Disease compare to traditional centralized systems, in terms of accuracy, recall, precision, and f-score?

\section{SOLUTION}
Expert systems are generally composed of knowledge base, inference engine, user and user interface, and the interaction between these subdivisions makes it an expert system. The collected fact and rules are stored in the knowledge base, which is connected to the inference engine to analyze the rule to deduce another set of rules or facts (see Figure 1).

\begin{figure}[ht]
    \centering
    \includegraphics[width=0.45\textwidth]{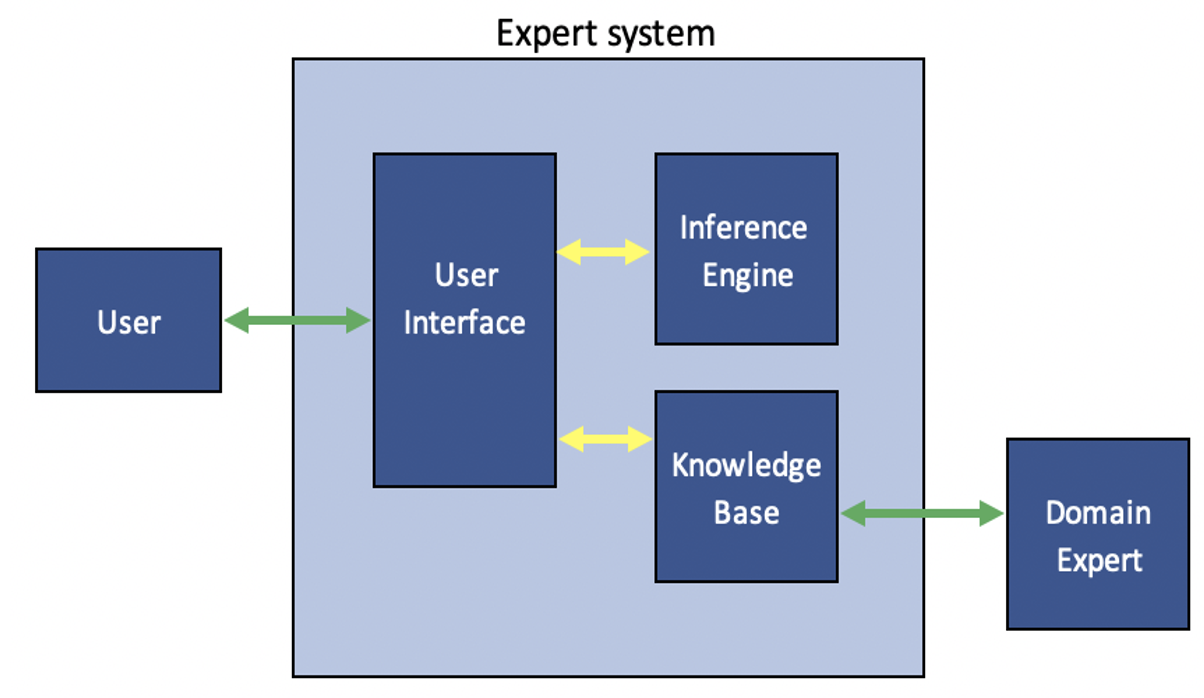}
    \caption{Block diagram of the analysis process}
  \end{figure}

\subsection{Why Could Decentralized Expert System Outperform Centralized Expert System?}
Apart from the benefits of decentralized data collection via the patients, decentralized expert system (ES) could outperform centralized ES. Some scenarios may involve additional complexities, such as variations in data quality, data distribution among sources, and communication overhead in decentralized setups. Empirical validation on datasets and comprehensive experimentation would be essential to draw concrete conclusions about the performance comparison between decentralized and centralized ES models.
To mathematically prove that decentralized ES provides better performance, we need to establish some assumptions and set up a rigorous framework for comparison. Let's outline the steps for the proof:

\begin{itemize}
    \item \textbf{Define Performance Metrics:} We need to establish specific performance metrics that measure the effectiveness of expert systems. For example, in the context of prediction models for AD, we could consider metrics such as accuracy, precision, recall, F1-score, or area under the receiver operating characteristic curve (AUC-ROC).
    
    \item \textbf{Formulate the Centralized ES Model:} Assume we have a centralized ES model that is trained using a centralized dataset containing MRI images from various healthcare institutions. We denote the performance of this model as \(P_{\text{centralized}}\).
    
    \item \textbf{Formulate the Decentralized ES Model:} Now, let's consider a decentralized ES model that is trained using data from multiple sources. The data is not pooled in a central location but remains distributed at each source. The performance of this model is denoted as \(P_{\text{decentralized}}\).
    
    \item \textbf{Derive a Theoretical Bound:} We need to establish a theoretical bound that represents the maximum achievable performance of a centralized ES model, given the dataset it has access to. This bound, denoted as \(P_{\text{bound}}\), acts as a theoretical benchmark for comparison.
    
    \item \textbf{Prove Inequality:} The mathematical proof will involve showing that \(P_{\text{decentralized}} \geq P_{\text{centralized}} > P_{\text{bound}}\). In other words, the decentralized ES model's performance is greater than or equal to the centralized ES model's performance, which, in turn, is greater than the theoretical bound representing the maximum achievable performance by a centralized model.
    
    \item \textbf{Account for Data Diversity:} In the proof, we should consider the potential benefits of data diversity in a decentralized ES setting. By training on data from various sources, the decentralized model can capture a more comprehensive representation of AD patterns, leading to better generalization and improved performance.
    
    \item \textbf{Consider Algorithmic Enhancements:} The proof should also consider the potential for algorithmic enhancements in the decentralized setting. With data from multiple sources, researchers can explore more sophisticated algorithms that leverage diverse data inputs, leading to better feature extraction and model optimization.
    
    \item \textbf{Account for Communication Overhead:} It's important to acknowledge any communication overhead associated with the decentralized setup. While decentralized models have the potential for better performance, communication delays or constraints may impact the overall efficiency.
\end{itemize}

Proving the superiority of decentralized ES over centralized ES using mathematical formulas is challenging due to the complexity and variability of real-world scenarios. However, we can present a theoretical argument that supports the potential benefits of decentralized ES in certain cases.

Let's consider a simplified scenario for binary classification tasks, where the goal is to predict whether an individual has AD (positive class) or not (negative class) based on MRI images. We will focus on the accuracy metric, but the argument can be extended to other performance metrics as well. Assumptions:

\begin{itemize}
    \item \textbf{Centralized ES:} A centralized ES model is trained on a dataset containing \(N_c\) samples from a single institution.
    
    \item \textbf{Decentralized ES:} A decentralized ES model is trained on the same dataset but is distributed across \(K\) institutions, each contributing \(N_d\) samples (such that \(N_d \times K = N_c\)).
\end{itemize}

\textbf{Performance Metric:}
\begin{itemize}
    \item Let \(P_{\text{centralized}}\) represent the accuracy of the centralized ES model.
    
    \item Let \(P_{\text{decentralized}}\) represent the accuracy of the decentralized ES model.
\end{itemize}

\textbf{Theoretical Bound:}
\begin{itemize}
    \item Let \(P_{\text{bound}}\) represent the theoretical upper bound on accuracy when the model is trained on the entire dataset, i.e., \(N_c\) samples.
\end{itemize}

\textbf{Mathematical Representation:}
\begin{itemize}
    \item \textbf{Centralized ES:} The accuracy of the centralized ES model can be expressed as follows: \(P_{\text{centralized}} = \frac{\text{Number of Correct Predictions}}{N_c}\)
    
    \item \textbf{Decentralized ES:} The accuracy of the decentralized ES model can be expressed as follows: \(P_{\text{decentralized}} = \frac{\text{Sum of Correct Predictions from Each Institution}}{N_c}\)
    
    \item \textbf{Theoretical Bound:} The theoretical bound on accuracy can be expressed as follows: \(P_{\text{bound}} = \frac{\text{Number of Correct Predictions When Trained on All } N_c \text{ Samples}}{N_c}\)
\end{itemize}

Now, to prove that decentralized ES provides better performance (\(P_{\text{decentralized}} \geq P_{\text{centralized}} > P_{\text{bound}}\)), we need to show two things:

\begin{itemize}
    \item \(P_{\text{decentralized}} \geq P_{\text{bound}}\): The decentralized ES model is trained on data from multiple sources, capturing data diversity and enabling better generalization. Hence, it has the potential to achieve an accuracy (\(P_{\text{decentralized}}\)) that is at least as good as the theoretical bound (\(P_{\text{bound}}\)).
    
    \item \(P_{\text{centralized}} < P_{\text{bound}}\): The centralized ES model is trained on a smaller dataset from a single source/institution, limiting its ability to capture the full data diversity present in the entire dataset. Thus, \(P_{\text{centralized}}\) is likely to be lower than the theoretical bound (\(P_{\text{bound}}\)).
\end{itemize}

\subsection{AI Model Predicting Early Stage AD}
The expert systems are being developed using various techniques, which are mostly used to assist medical practitioners in diagnosis. In this study, we need to train the AI model (Figure 2) via the data that we have obtained from the Alzheimer’s Disease Neuroimaging Initiative (ADNI) database (http://adni.loni.usc.edu) [12], a public-private partnership launched in 2003 by Michael Weiner, MD. Our proposed framework consists of processing steps: feature extraction, feature selection, and classification. We examined different feature selection methods to choose an optimal subset of features, maximizing the accuracy of classification between cognitively normal (CN), individuals with significant memory concern (SMC) and mild cognitive impairment (MCI) patients. The subjects were randomly split into training and testing datasets, and Random Forest classifier was trained using the training dataset. Finally, the testing dataset was passed to the trained classifier to measure the performance.

\begin{figure*}[ht] 
    \centering
    \includegraphics[width=1\textwidth]{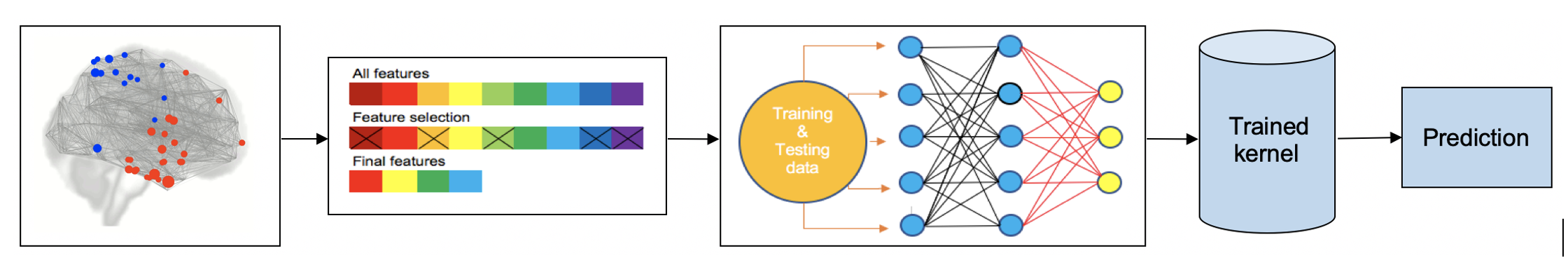}
    \caption{AI classification model}
\end{figure*}

\begin{table}[ht]
    \centering
    \small 
    \begin{tabular}{p{8cm}}
    \hline
    \textnormal{Algorithm 1: this algorithm is designed for early-stage AD detection.} \\
    \hline
    1. \textbf{Load Data from a CSV File:} \\
        - Import data from a CSV file using `load\_data()`. \\
    2. \textbf{Split Data by Labels:} \\
        - Separate data into groups based on labels with `split\_data\_by\_labels()`. \\
    3. \textbf{Split Data into Training and Testing Sets:} \\
        - Divide data into training and testing sets using `split\_data\_train\_test()`. \\
    4. \textbf{Scale Features Using StandardScaler:} \\
        - Normalize features using `scale\_features()`. \\
    5. \textbf{Create a Random Forest Classifier:} \\
        - Initialize a Random Forest classifier with `create\_random\_forest\_classifier()`. \\
    6. \textbf{Initialize a List for Accuracy Results:} \\
        - Create a list to store accuracy results: `accuracy\_results = []`. \\
    7. \textbf{Loop Over the Range of Features:} \\
        - Iterate over feature numbers with `for k in range():`. \\
    8. \textbf{Instantiate a Logistic Regression Classifier:} \\
        - Use `create\_logistic\_regression\_classifier()` within the loop. \\
    9. \textbf{Sequential Forward Selection (SFS):} \\
        - Perform SFS to find the best features using a nested loop `for i in range(k):`. \\
    10. \textbf{Train and Evaluate the Classifier:} \\
        - Use `train\_classifier()` and `evaluate\_classifier()`\\ 
        to train and assess the classifier. \\
    11. \textbf{Store Accuracy for Current k Value:} \\
        - Save the classifier's accuracy for the current `k` with `store\_accuracy()`. \\
    12. \textbf{Plot Accuracy:} \\
        - Visualize accuracy results using `plot\_accuracy()`. \\
    \hline
    \end{tabular}
\end{table}

We have used data for 561 subjects total, among those, 231 SMC, 259 CN, and 71 MCI patients. The feature selection algorithms were applied to the graph features (degree centrality for each ROI) to select the most discriminating features for the classification of MCI, SMC, and CN subjects. The Sequential Forward Selection feature selection algorithm and the Random Forest classifier resulted in a satisfying performance with accuracy of more than 92\% as shown in Figure 3.

\begin{figure}
    \centering
    \includegraphics[width=0.45\textwidth]{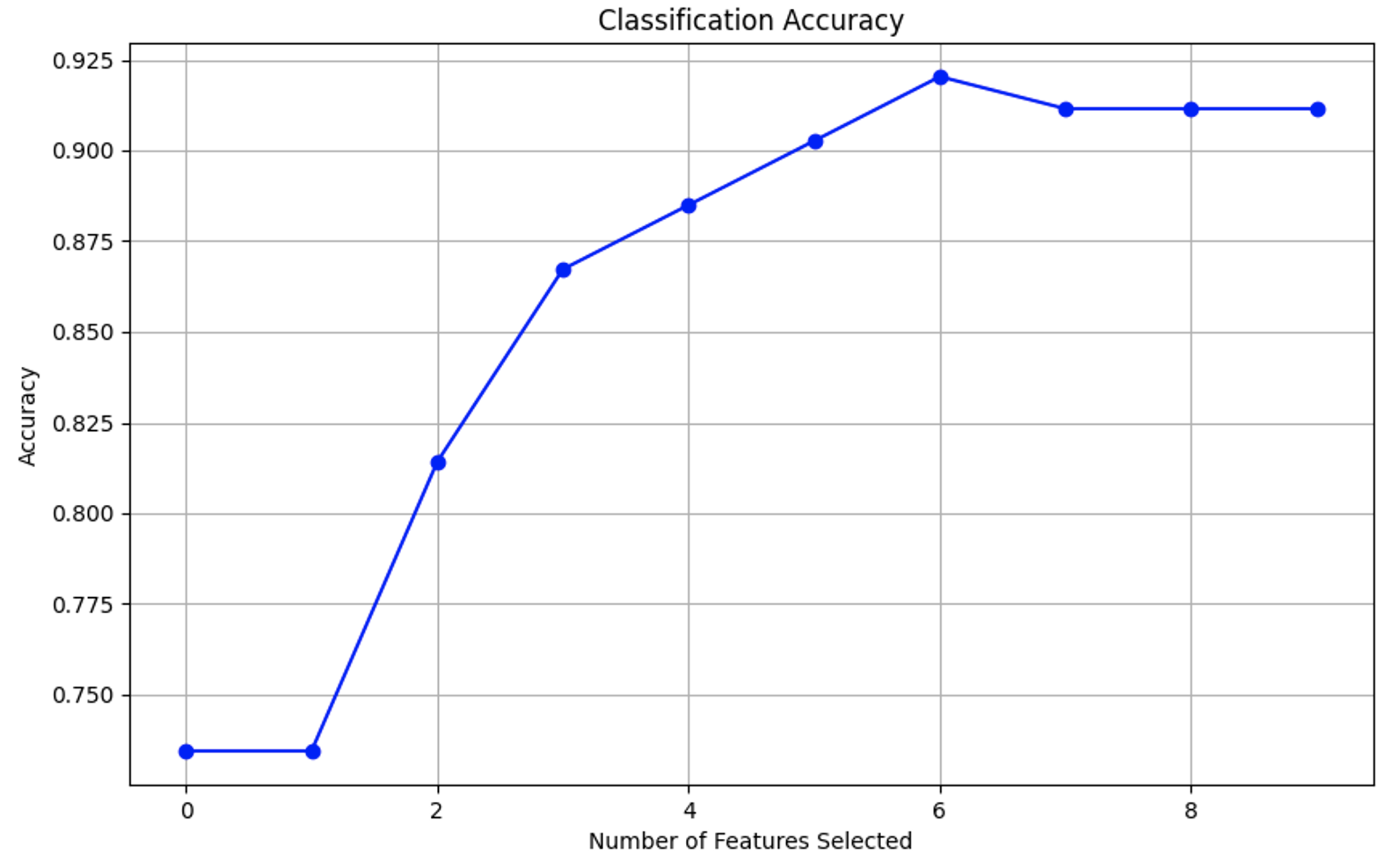}
    \caption{Classification accuracy of AI model.}
  \end{figure}

The graph features were obtained by applying graph theory analysis on rs-fMRI images. The pre-processing, network modeling for graph feature extraction is done via FSL library. The patients can therefore input their MRI images via the provided App, and the FSL library processes, and generates the brain connectivity matrix. From longitudinal measures, patients are labeled as non-convertors and convertors fulfilling the criteria for Prodromal AD’s continuum according to Jack et al. (2018) [13]. At this stage, we have just trained the AI model with publicly available ADNI data, but once the decentralized longitudinal data is obtained directly via the patients, the classification model is retrained and aggregated based on these data. We will explain this further in the implementation section.

\subsection{Anomaly Detection}
Performing an exhaustive check of the correctness and quality of MRI images and biological information directly on-chain is not practical due to the computational complexity and cost constraints of blockchain platforms. Instead, such checks are typically performed off-chain, and the blockchain is used to record the results securely.

A practical example of a smart contract that allows patients to submit their data along with a brief initial evaluation is given in Listing 1. The contract stores this data on-chain and allows patients to verify and timestamp their submissions. Note that this contract primarily serves as a ledger for the data and initial evaluation results, and more comprehensive checks should be performed off-chain by the application (DApp) before submitting data to the blockchain.

\begin{table}[htbp]
    \centering
    \small 
    \label{algorithm1}
    \begin{tabular}{p{8cm}}
        \hline
        \textnormal{Listing 1:} A smart contract that allows patients to submit data and verify. \\
        \hline
        // SPDX-License-Identifier: MIT \\
        pragma solidity $^\wedge$0.8.0;\\
        
        contract MedicalDataSubmission \{ \\
        \qquad struct PatientData \{ \\
        \qquad\qquad address patient; \\
        \qquad\qquad string biologicalInfo; \\
        \qquad\qquad string evaluation; \\
        \qquad\qquad uint256 timestamp; \\
        \qquad\qquad bool isVerified; \\
        \qquad \} \\
        \qquad PatientData[] public submissions; \\
        \qquad event DataSubmitted(uint256 indexed submissionId, address indexed patient, string biologicalInfo, string evaluation, uint256 timestamp); \\
        
        \qquad function submitData(string memory biologicalInfo, string memory evaluation) external \{ \\
        \qquad\qquad require(bytes(biologicalInfo).length $>$ 0, "Biological information cannot be empty."); \\
        \qquad\qquad require(bytes(evaluation).length $>$ 0, "Evaluation cannot be empty."); \\
        \qquad\qquad submissions.push(PatientData(msg.sender, biologicalInfo, evaluation, block.timestamp, false)); \\
        \qquad\qquad uint256 submissionId = submissions.length - 1; \\
        \qquad\qquad emit DataSubmitted(submissionId, msg.sender, biologicalInfo, evaluation, block.timestamp); \\
        \qquad \} \\
        
        \qquad function verifySubmission(uint256 submissionId) external \{ \\
        \qquad\qquad require(submissionId $<$ submissions.length, "Submission does not exist."); \\
        \qquad\qquad PatientData storage submission = submissions[submissionId]; \\
        \qquad\qquad require(msg.sender == submission.patient, "Only the patient can verify the submission."); \\
        \qquad\qquad require(!submission.isVerified, "Submission is already verified."); \\
        \qquad\qquad // Implement additional verification steps as needed \\
        \qquad\qquad submission.isVerified = true; \\
        \qquad \} \\
        \} 
    \end{tabular}
\end{table}

In this contract, the "submitCertificate" function allows patients to submit the results of the off-chain anomaly detection process. The "verifyCertificate" function allows patients to verify their certificates.
One can implement additional verification steps in the "verifyCertificate" function as needed.

Implementing off-chain anomaly detection for both biological information and MRI images is a complex task that requires specialized libraries and tools. 
To implement a smart certificate for anomaly detection on the client side of a medical data sharing platform, we would use off-chain data analysis techniques since performing anomaly detection directly on-chain can be expensive and inefficient due to the trade-off between performance and security.

\textbf{Data Collection:} Patients provide their biological information and MRI images along with timestamps to the application.

\textbf{Off-Chain Anomaly Detection:} Implement advanced anomaly detection algorithms off-chain within the App. For MRI images, one might use computer vision techniques, and for biological information, statistical or machine learning methods can be applied to detect anomalies. These algorithms should thoroughly evaluate the correctness and quality of the data.

\textbf{Smart Certificate Creation:} After off-chain anomaly detection, create a detailed smart certificate within the App to include:
\begin{itemize}
    \item Anomaly type (e.g., incorrect data, bad images, etc.).
    
    \item Timestamp.
    
    \item Metadata about the data and the anomaly.
    
    \item Any relevant context or notes about the anomaly.
\end{itemize}

\textbf{Blockchain Interaction:} Use a smart contract on the blockchain to securely store and verify the smart certificates generated within the App. The smart contract records the results of the anomaly detection process, providing an immutable and auditable record.

\subsubsection{Off-chain anomaly detection for biological information}
For biological information, anomaly detection can involve statistical methods or machine learning techniques, depending on the nature and structure of the data. Here in Listing 2, we provide an approach using Python and the popular scikit-learn library:
In this example, we perform the following steps:
\begin{itemize}
\item Load biological data.
\item Select the relevant features for anomaly detection.
\item Apply feature scaling using StandardScaler.
\item Reduce dimensionality using PCA.
\item Choose an anomaly detection model (Isolation Forest, or) and fit it to the reduced data.
\item Predict anomaly scores for each data point.
\item Define a threshold for anomaly detection (experiment with different thresholds).
\item Identify anomalies based on the threshold.
\item Perform further processing or reporting of detected anomalies.
\end{itemize}

\begin{table}[htbp]
    \centering
    \footnotesize 
    \begin{tabular}{p{8cm}}
    \hline
    \textnormal{Listing 2:} Anomaly detection for biological information using Isolation Forest. \\
    \hline
    \begin{verbatim}
import numpy as np
from sklearn.preprocessing import StandardScaler
from sklearn.decomposition import PCA
from sklearn.covariance import EllipticEnvelope
from sklearn.ensemble import IsolationForest
from sklearn.svm import OneClassSVM

# Load your biological data 
biological_data = load_biological_data()

# Select the relevant features for anomaly 
detection
selected_features = ['feature1', 'feature2', 
'feature3']
X = biological_data[selected_features]

# Apply feature scaling
scaler = StandardScaler()
X_scaled = scaler.fit_transform(X)

# Apply dimensionality reduction using PCA
pca = PCA(n_components=2)
X_pca = pca.fit_transform(X_scaled)

# Choose an anomaly detection model
model = IsolationForest(contamination=0.05)
model.fit(X_pca)

# Predict anomalies
anomaly_scores = model.decision_function(X_pca)

# Define a threshold for anomaly detection
threshold = -0.3  # Adjust as needed

# Identify anomalies
anomalies = biological_data[anomaly_scores 
< threshold]

# Further processing or reporting of anomalies
for index, row in anomalies.iterrows():
    print(f"Anomaly detected for sample {index}
    :")
    print(row)

# experiment with different models
# (Elliptic Envelope, One-Class SVM, etc.)
# and fine-tune parameters for better anomaly 
detection performance.            
    \end{verbatim}
    \end{tabular}
\end{table}

\subsubsection{Off-chain anomaly detection for MRI images}
Detecting anomalies in MRI images typically involves computer vision techniques and deep learning models. One might consider using popular deep learning libraries like TensorFlow or PyTorch. Here in Listing 3, we provide an approach using a pre-trained model.
This approach allows to detect anomalies in MRI images based on how well the autoencoder can reproduce the input image. Anomalies will typically result in higher MSE values compared to normal images. One might need to fine-tune the threshold based on the dataset and requirements.
In this code:
\begin{itemize}
\item Load a pre-trained autoencoder model (both encoder and decoder parts). Autoencoders learn to encode data efficiently and are often used for anomaly detection because they can reproduce normal data accurately.
\item Load an MRI image (replace 'mri\_image.png') and preprocess it.
\item Encode the image using the autoencoder's encoder part, then decode it to get a reconstructed image.
\item Calculate the Mean Squared Error (MSE) between the original and reconstructed images. This measures how well the model can reproduce the input.
\item Set a threshold for the MSE, above which an anomaly is detected.
\end{itemize}

\begin{table}[htbp]
    \centering
    \footnotesize 
    \label{algorithm2}
    \begin{tabular}{p{8cm}}
    \hline
    \textnormal{Listing 3:} Off-chain anomaly detection for MRI images. \\
    \hline
    \begin{verbatim}
import tensorflow as tf
import numpy as np
from PIL import Image

# Load pre-trained autoencoder model
autoencoder = tf.keras.models.load_model
('autoencoder_model.h5')

# Load an MRI image
image = Image.open('mri_image.png')
# Normalize image data
image = np.array(image) / 255.0  

# Preprocess the image for model input
# Resize to the model's input size
input_image = tf.image.resize(image, (224, 224)) 
# Add batch dimension 
input_image = np.expand_dims(input
_image, axis=0)  

# Encode the image using the autoencoder
encoded_image = autoencoder.encoder(input_image)
.numpy()

# Calculate reconstruction loss
reconstructed_image = autoencoder(input
_image).numpy()
mse = np.mean(np.square(input_image - 
reconstructed_image))

# Define a threshold for anomaly detection
threshold = 0.01  # Adjust as needed

if mse > threshold:
    print("Anomaly detected in MRI image.")
else:
    print("No anomaly detected in MRI image.")    
    \end{verbatim}      
    \end{tabular}
\end{table}

\section{SYSTEM DEVELOPMENT}
In terms of System Development status, and all the components required according to the class diagram in Figure 4, we have already developed and tested the user-interface application (see https://github.com/stefankam/predprodalzheimer); this application is based on FSL library which also performs MRI data processing, and will be discussed further in the application infrastructure.\\
The underlying blockchain technology for decentralized data sharing (Behfar et al. [14]) has already been developed (see https://github.com/stefankam/Blockchain-based-data-sharing), which is based on hyperldger fabaric technology for on-chain , and IPFS for off-chain data sharing as pilot project. Currntly We explore alternative solutions such as zero-knowledge and optimistic rollups.\\
The ML models for early AD detection have also been developed, trained, and tested using public dataset ADNI, mentioned in "AI Model Predicting Prodromal AD". Regarding continuous learning as explained in Algorithm 1, If the Random Forest classifier doesn't support this well, we consider introducing online learning or incremental learning, as the model is supposed to update or learn from new data.
Figure 4 illustrates the class diagram of the whole system, where each class is defined below:\\

\begin{figure}[ht]
    \centering
    \includegraphics[width=0.45\textwidth]{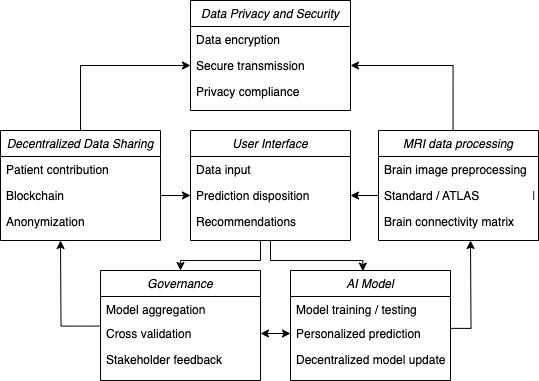}
    \caption{Class diagram}
  \end{figure}

  \textbf{User Interface:} This is the primary interface for patients to input their anonymous biological info and MRI images and receive prediction deposition and recommendations; this includes approaching a specialist for further and more certain diagnostics. It's the front-end through which users interact with the system.\\

  \textbf{Data Security and Privacy:} This component would be responsible for ensuring that patient data, particularly sensitive MRI images, are handled securely and in compliance with privacy regulations. It interfaces with both the User Interface (to ensure data is securely transmitted) and the Decentralized Data Sharing component (to ensure data is securely stored and shared).\\
  
  \textbf{MRI Data Processing:} This component processes the MRI images provided by patients through the User Interface. It uses tools like the FSL library for generating brain connectivity matrices, which are crucial for AD prediction. This processed data would then be fed into the AI Model for analysis and prediction/classification.\\
  
  \textbf{Decentralized Data Sharing:} This component is responsible for the secure and anonymous management of patient data within the decentralized network. It ensures that data from various patients is aggregated without compromising individual privacy. This component is crucial for maintaining the decentralized nature of the system, allowing for data sharing and aggregation across different nodes.\\
  
  \textbf{AI Model:} The AI model, possibly a Random Forest classifier or similar, is trained on the aggregated brain connectivity matrices and possibly other patient data. It's responsible for early-stage AD detection, and making predictions about the progression to AD. This model will continuously learn from new patient data, improving its accuracy and adaptability over time.\\
  
  \textbf{Governance:} This component oversees the overall functioning of the system, ensuring that all parts work together cohesively, aggregating model, adhere to set standards and regulations. It will also be involved in updating the system, incorporating patient feedback, and ensuring the system's continuous improvement.

\subsection{Implementation of Decentralized Expert System}
The goal is to implement a decentralized expert system to detect prodromal or early stage AD by leveraging the collective data of a network of decentralized nodes. Such a system can involve a distributed network, where patients input anonymous biological information and medical test results.
To implement the described decentralized expert system for early-stage AD detection, one needs to integrate several components and consider the role of patients in the system. The overview of the implementation steps is given in Algorithm 2.
Regarding the role of patients in the decentralized system:\\

\begin{table}
    \centering
    \small 
    \begin{tabular}{p{8cm}}
    \hline
    \textnormal{Algorithm 2:} Decentralized expert system for early-stage AD detection. \\
    \hline
    1. \textbf{Data Collection and Brain Connectivity Matrix Generation:} \\
       - Patients use the application to input their MRI images. \\
       - The FSL library processes the MRI images and generates a brain connectivity matrix. \\
    2. \textbf{Decentralized Data Sharing and Model Training:} \\
       - The generated brain connectivity matrices are shared within a decentralized network. \\
       - Patients' data may be stored in a privacy-preserving manner, ensuring that the network adheres to privacy regulations. \\
       - The decentralized system aggregates brain connectivity matrices from various patients. \\
    3. \textbf{AI Model Training and Personalization:} \\
       - The AI model, trained initially on a public dataset, can be further fine-tuned and personalized using the aggregated brain connectivity matrices from the patients. \\
       - The model continuously learns from new patient data, improving its accuracy and adaptability. \\
    4. \textbf{Prediction and Longitudinal Monitoring:} \\
       - Patients' longitudinal data is used to monitor disease progression over time. \\
       - The trained AI model predicts the transformation to AD based on the input brain connectivity matrix and patient's longitudinal data. \\
    5. \textbf{Feedback Loop and Model Updates:} \\
       - Patient feedback and outcomes is collected to improve the model's performance and refine the prediction process. Regular model updates based on the latest data and patient feedback ensure that the AI model stays up-to-date and personalized. \\
    \hline
    \end{tabular}
\end{table}

\textbf{Patients as Users:} Patients primarily interact with the system as users. They provide input data (Biological info and MRI images), receive predictions, and have access to longitudinal monitoring and recommendations. They are not typically considered nodes in the decentralized network but are essential stakeholders.

\textbf{Decentralized Nodes:} The decentralized network consists of nodes that share and process data. These nodes may include servers, data repositories, AI model components, and other network participants. Patient data can be managed securely and anonymously within the network, but patients themselves are not nodes.

\subsubsection{Application infrastructure}
For MRI preprocessing and network modeling, via App, we use FSL library which is extremely powerful when it comes to applying and automating workflow since it can unify some of the most crucial steps into one pipeline only and thereby facilitate the entire workflow. The scripts from the FSL library can be run on either Linux or macOS. FSL unifies some of the most crucial steps into one pipeline only and thereby facilitate the entire workflow, see https://github.com/stefankam/predprodalzheimer, also note that to use FSLNets either Octave or MATLAB must be running. Putting all the steps together, here is what a workflow could look like:
\begin{itemize}
    \item Skull stripping – using BET 
    
    \item Preprocessing – using the modules indicated at the preprocessing step
    
    \item Node definition – using MELODIC and Octave
    
    \item Generating connectivity matrix – using FSLNets
\end{itemize}

The backend of this application will not only mange the project’s APIs, from frontend to backend to database and vice-versa, but also manage the interaction with FSL and Octave. The latter is indispensable for the creation of the Brain Connectivity Matrix (BCM). As indicated in Figure 5, Schlappinger (2023) [15], all user requests always pass via the server’s API-service first, and are dispatched to the corresponding service. When the user tries to log in, the log-in data is sent to the backends’ API service, then sends it to the corresponding application service, which in this case would be the authentication service. It handles the transferred data and asks for identification by sending requests to the database. The database response is sent to the application service, and the response back to the API. 
With the definition of the expert system, the web application does preprocessing on the subjects to finally output the brain connectivity matrix that is available immediately after processing. 

\begin{figure}
    \centering
    \includegraphics[width=0.45\textwidth]{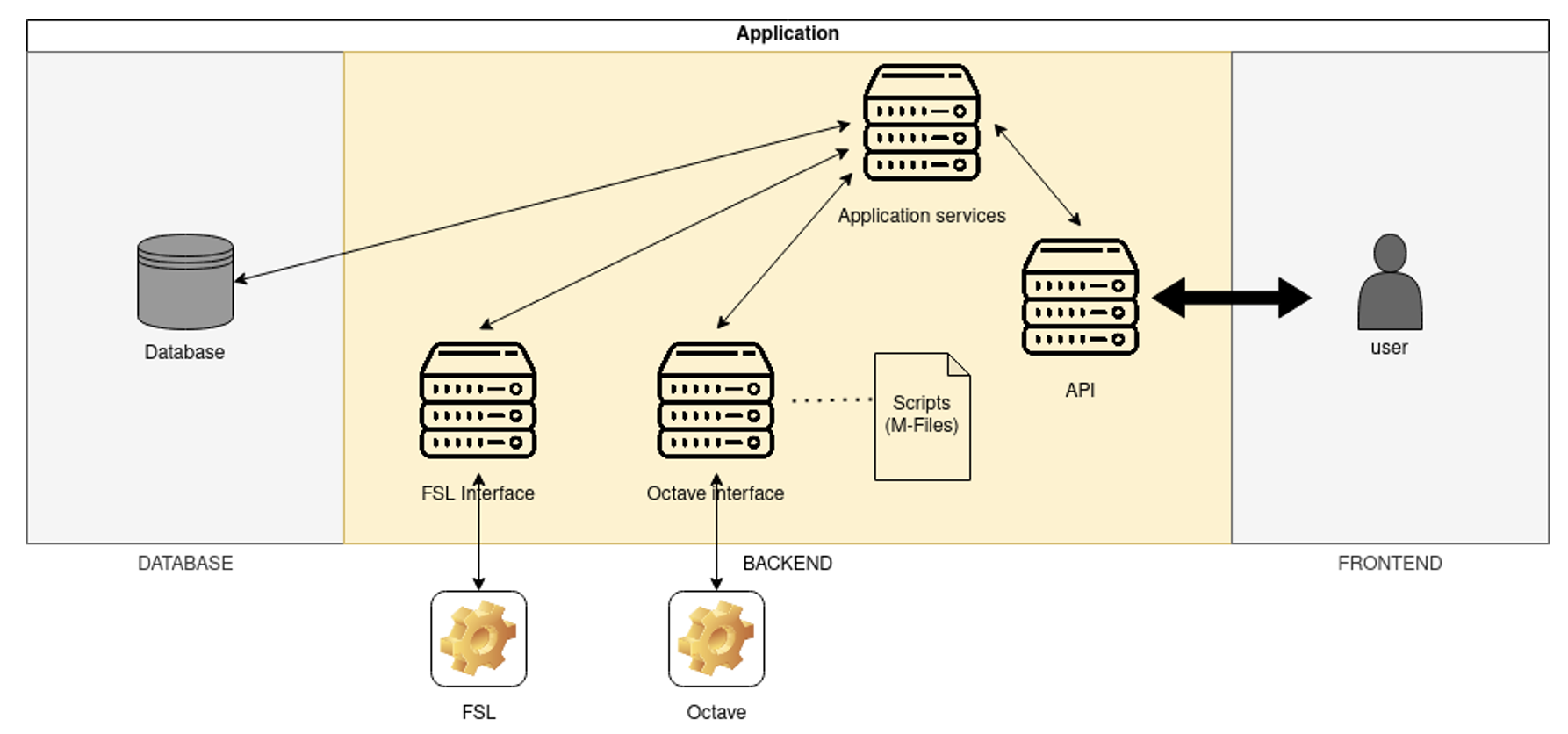}
    \caption{backend-frontend infrastructure diagram.}
  \end{figure}

\subsubsection{Application security}
Ensuring the security and privacy of medical data is of paramount importance in our system development. We implement a comprehensive set of measures to safeguard sensitive information, maintain data integrity, and comply with privacy regulations.\\

\textbf{Data Encryption}\\
End-to-End Encryption: All medical data, including biological information and MRI images, undergo end-to-end encryption using industry-standard encryption algorithms. This means that data is encrypted at its source (on the patient's side) and remains encrypted during transmission and storage within our system. Even if an unauthorized entity intercepts the data, it remains indecipherable without the encryption keys.\\

AES Encryption: We employ the Advanced Encryption Standard (AES) for data encryption. AES is a widely recognized and robust encryption algorithm known for its security and performance. It ensures that patient data is protected from unauthorized access.\\

\textbf{Secure Transmission}\\
HTTPS: We utilize the Hypertext Transfer Protocol Secure (HTTPS) for web-based data transmission. HTTPS is a secure communication protocol that combines the standard HTTP with encryption using Transport Layer Security (TLS) or Secure Sockets Layer (SSL) protocols. This encryption layer ensures that data exchanged between the client and our system is shielded from eavesdropping and tampering during transit.\\

Blockchain Technology: Our system leverages blockchain technology to enhance the security of data sharing. Blockchain, with its decentralized and immutable ledger, provides an additional layer of protection. Each data transaction is recorded on the blockchain, and once added, it cannot be altered. This ensures transparent and secure data sharing among authorized parties.\\

\textbf{Privacy Compliance}\\
Access Control: Access control mechanisms are in place to restrict data access to only authorized healthcare professionals and patients. Role-based access control ensures that individuals can only access the data that is relevant to their responsibilities. Patients have control over who can access their data, granting consent for sharing, and revoking access as needed.\\

HIPAA and GDPR Compliance: Our system adheres to the Health Insurance Portability and Accountability Act (HIPAA) and the General Data Protection Regulation (GDPR), in addition to local data protection laws. These compliance measures provide a legal framework for the secure handling of patient data, including rules for data access, storage, and sharing.\\

Regular Audits and Privacy Impact Assessments: To maintain compliance, we need to conduct regular system audits and privacy impact assessments. These evaluations help us identify and rectify potential privacy issues and vulnerabilities in our system. They also ensure that we remain aligned with the latest data protection regulations.\\

Even if patient data is anonymized, it's often advisable and may be legally required to comply with many of the security and privacy measures mentioned above. Anonymization can reduce the risk associated with the disclosure of sensitive information, but it doesn't necessarily exempt a system from all privacy regulations or security best practices.

\section{CONCLUSION}
Overall, a decentralized expert system for early-stage AD detection can leverage the collective data and intelligence to provide accurate and timely predictions, while also allowing for continuous improvement over time. Our expert system serves as a model tool that collects patients’ data in a decentralized way via our FSL-built application. FSL using Octave creates brain connectivity matrices and pushes to the AI engine. Our trained model uses Sequential Forward Selection feature selection algorithm and the Random Forest classifier resulting in accuracy of more than 92\%; the classification model is retrained based on obtained longitudinal data. This not only helps individuals to detect early-stage AD in time, but also helps clinics and hospitals who are willing to use this software to effectively monitor the patients and predict their progression with less ambiguity.

\end{document}